\documentclass[prd,aps,preprintnumbers,nofootinbib,twocolumn]{revtex4}
\usepackage{epsfig}
\usepackage{amsmath}
\usepackage{hyperref}

\begin{document}

\preprint{Cavendish-HEP-18/11}

\renewcommand{\thefigure}{\arabic{figure}}

\title{Two-loop five-point massless QCD amplitudes within the integration-by-parts approach}

\author{Herschel A. Chawdhry}
\author{Matthew A. Lim}
\author{Alexander Mitov}
\affiliation{{\small Cavendish Laboratory, University of Cambridge, Cambridge CB3 0HE, United Kingdom}}

\date{\today}

\begin{abstract}
We solve the integration-by-parts (IBP) identities needed for the computation of any planar two-loop five-point massless amplitude in QCD. We also derive some new results for the most complicated non-planar topology with irreducible numerators of power as high as six. We do this by applying a new strategy for solving the IBP identities which scales better for problems with a large number of scales and/or master integrals. Our results are a proof of principle that the remaining non-planar contributions for all two-loop five-point massless QCD amplitudes can be computed in analytic form.
\end{abstract}
\maketitle

\section{Introduction\label{sec:intro}}

Gauge theories, whose predictivity is dependent on calculations of scattering amplitudes at higher perturbative orders, have been hugely successful in describing natural phenomena. The focus of this work is on perturbative Quantum Chromodynamics (QCD) which can be viewed as a prototype for a generic unbroken non-supersymmetric gauge theory. QCD is also special because multiloop QCD amplitudes are the backbone of theoretical predictions for precision collider observables like the ones measured at the Large Hadron Collider.

At present, calculations in massless QCD are possible at four (partially five) loops \cite{Herzog:2017ohr,Luthe:2017ttc,Baikov:2017ujl,Ruijl:2017cxj,Georgoudis:2018olj} for self-energy diagrams and three (partially four) loops \cite{Gehrmann:2010ue,Boels:2015yna,Henn:2016men,vonManteuffel:2016xki,Lee:2016ixa,Lee:2017mip} for vertex-like processes. Four-point massless amplitudes are fully known  through two loops (partially three) \cite{Anastasiou:2000kg,Anastasiou:2001sv,Glover:2001af,Abreu:2017xsl,Henn:2013fah,DiVita:2014pza} while five-point ones are fully known at one loop (partly two
\cite{Badger:2013gxa,Badger:2015lda,Gehrmann:2015bfy,Dunbar:2016aux,Dunbar:2017nfy,Badger:2017jhb,Abreu:2017hqn,Boels:2018nrr,Boehm:2018fpv}). The ultimate goal of this work is to extend the two-loop frontier to the complete set of five-point massless QCD amplitudes.
  
The integration-by-parts identities (IBP) approach \cite{Tkachov:1981wb,Chetyrkin:1981qh} has been the method of choice for computing multiloop QCD amplitudes. The method has produced countless results; see some recent reviews \cite{Grozin:2011mt,Kotikov:2018wxe}. The way the IBP approach works is rather simple. A generic squared or suitably decomposed multiloop UV-unrenormalized amplitude can be written as
\begin{equation}
M = \sum_{i=1}^N f_iI_i\,.
\label{eq:M}
\end{equation}

The above expression follows from a straightforward application of Feynman rules applied to the process at hand and, if appropriate, after summation over spin and/or color. Throughout this work we assume that all divergences are regulated by working in $d=4-2\epsilon$ dimensions. The coefficients $f_i$ are rational functions of kinematic invariants and the space-time dimension $d$, and $I_i$ are scalar Feynman integrals. The number $N$ of such integrals tends to be very large and grows quickly with the number of loops, legs and/or parameters in the problem. 

The IBP approach makes it possible to express the Feynman integrals $I_i$ appearing in eq.~(\ref{eq:M}) as linear combinations of a small number of Feynman integrals $\hat I_m$
\begin{equation}
I_i = \sum_{m=1}^{\hat N} c_{i,m}\hat I_m\,.
\label{eq:I_i}
\end{equation}
The integrals $\hat I_m$ are known as master integrals (or simply masters) and the coefficients $c_{i,m}$ are rational functions of the kinematic invariants and the space-time dimension.

The utility of the IBP approach stems from the fact that $\hat N\ll N$. For example, for the problem we consider in the present paper, $\hat N\sim{\cal O}(10^2)$ while $N\sim{\cal O}(10^4)$. 

Finally, substituting eq.~(\ref{eq:I_i}) in eq.~(\ref{eq:M}) one gets the desired minimal form for the amplitude in eq.~(\ref{eq:M})
\begin{equation}
M = \sum_{m=1}^{\hat N} \hat c_m \hat I_m\,,~{\rm with}~~~ \hat c_m=\sum_{i=1}^N c_{i,m} f_i \,.
\label{eq:Mhat}
\end{equation}

The evaluation of the bare amplitude $M$ consists of two steps: first, solve the IBP equations by deriving the required set of coefficients $c_{i,m}$ appearing in eqs.~(\ref{eq:I_i},\ref{eq:Mhat}) and, second, evaluate the master integrals $\hat I_m$. 

The subject of this work is the calculation of the coefficients $c_{i,m}$. We note that they are process-independent in the sense that they are the same for every massless two-loop five-point amplitude. Their universality is one of the advantages of the IBP method. All process-specific information is encoded into the coefficients $f_i$ which are comparatively easy to compute.

The master integrals $\hat I_m$ are also process independent. In the context of eq.~(\ref{eq:I_i}) they are interpreted as a basis of the $\hat N$-dimensional vector space $V\equiv\{I_i\}$ spanned by the infinite number of possible integrals $I_i$. We note that the choice of such a basis is not unique; moreover, it can happen that two or more master integrals are linearly related to each other when viewed as integrals. In the context of the IBP approach, however, the masters have to be treated as independent basis elements. In this work we will not be concerned with their evaluation since this is a separate, albeit not unrelated \cite{Kotikov:1990kg,Tarasov:1996br,Remiddi:1997ny,Gehrmann:1999as,Baikov:2010hf,Henn:2013pwa,Argeri:2007up,Henn:2014qga} problem. All planar master integrals relevant for the present work are known in analytic form \cite{Papadopoulos:2015jft}.

The solving of the IBP identities in the past 20 years or so has been based on the Laporta algorithm \cite{Laporta:2001dd}. Many computer implementations of this algorithm exist \cite{Anastasiou:2004vj,Smirnov:2008iw,Studerus:2009ye,vonManteuffel:2012np,Lee:2012cn,Smirnov:2014hma,Maierhoefer:2017hyi}. Although the results are analytic in the kinematic variables, they remain numeric in the powers of the propagators and cannot, therefore, be expected to solve problems of arbitrary complexity. As experience shows, the evaluation of the massless two-loop five-point QCD amplitudes is at the boundary of what is possible with the existing implementations of the Laporta approach. 

Many novel ideas for the solving of the IBP equations have been proposed in the recent past \cite{Lee:2008tj,Gluza:2010ws,Schabinger:2011dz,vonManteuffel:2014ixa,Ita:2015tya,Larsen:2015ped,Boels:2018nrr}. These new ideas and methods have made possible the evaluation of specific/planar all-gluon five-point amplitudes \cite{Badger:2013gxa,Badger:2015lda,Gehrmann:2015bfy,Dunbar:2016aux,Dunbar:2017nfy,Badger:2017jhb,Abreu:2017hqn} as well as some non-planar ones \cite{Boehm:2018fpv}. Ideas towards solving the IBP identities in abstract form have also been put forward \cite{Kosower:2018obg}.

In this work we explore a different strategy for solving the IBP identities. We demonstrate that supplementing this strategy with the standard Laporta algorithm is sufficient to solve the IBP identities needed to compute the complete set of planar two-loop five-point amplitudes in massless QCD (with quarks and/or gluons) in analytic form. We also present new non-trivial non-planar results. Based on our experience we expect that the non-planar contributions can be computed in analytic form with our strategy.

\section{Our strategy for solving the IBP identities}\label{sec:approach}

Our starting point is the assumption that the IBP system has a solution, i.e. {\it every} loop integral $I_i$ can be expressed through a set of basis master integrals as in eq.~(\ref{eq:I_i}) and that such a basis set of masters is known. 

The existence and construction of a finite basis of master integrals is an old problem \cite{Baikov:2005nv,Lee:2013hzt,Georgoudis:2016wff}. Here we take a pragmatic viewpoint which is informed by the observation that all problems known to us do possess such a finite basis. There are several ways to construct such a basis. For example, one could solve the IBP system over a restricted set of integrals and/or use numerical values for the kinematic invariants. In any case, finding a basis is not a bottleneck and we consider this step to be trivial. This is certainly true for the two-loop five-point massless amplitudes considered here, where we have easily identified the sets of masters for all topologies.

The index $i$ labeling the integral $I_i$ is a composite index. It is natural to express it through the powers of the propagators appearing in the corresponding integral. For example, for a generic two-loop integral we have
\begin{equation}
I_i \equiv I(n_1,\dots,n_P) = \int d^dk_1 d^dk_2 {1\over \Pi_1^{n_1}\dots \Pi_P^{n_P}}\,.
\label{eq:I_def}
\end{equation}
The functions $\Pi_n$ are the corresponding propagators which are bilinear functions of the loop and/or external momenta. Specific examples are given in sec.~\ref{sec:results}. Eq.~(\ref{eq:I_def}) can, of course, be generalized to any loop order in a completely straightforward way.

In this notation eq.~(\ref{eq:I_i}) now takes the form
\begin{equation}
I(n_1,\dots,n_P) = \sum_{m=1}^{\hat N} c_{m}(n_1,\dots,n_P)\hat I_m\,.
\label{eq:I_full}
\end{equation}
Just like the index $i$, the index $m$ is also a composite one and we will sometimes use its explicit form. 

To solve the IBP identities means that for any required integral $I_i$ one must derive the set of coefficients $c_{i,m}$ appearing in eqs.~(\ref{eq:I_i},\ref{eq:I_full}). In existing approaches for solving IBP identities, the full set of coefficients $c_{i,m}$ (for a given $i$) is derived simultaneously. In this work we pursue a different strategy for their solving where the projection of $I_i$ onto each master is derived {\it independently}. Put differently, we split the problem of solving the system of IBP equations into $\hat N$ independent problems, one for each of the $\hat N$ projections. 

This strategy is implemented in the following way: we apply the usual set of IBP identities to a modification of the space $V$ such that $\hat N-1$ of its elements (corresponding to all but one of the masters) are set to zero beforehand. For example, in order to derive the projection onto master $\hat I_1$ of any integral $I_i$, one first sets $\hat I_2=\hat I_3=\dots=\hat I_{\hat N}=0$ and then solves the IBP equations. This way, once the IBP system has been fully solved, one will have a solution that is of the following form
\begin{equation}
I(n_1,\dots,n_P) = c_{1}(n_1,\dots,n_P)\hat I_1\,,
\end{equation}
i.e. one will have derived the coefficients $c_{1}(n_1,\dots,n_P)$ which are the projection of the full solution onto the master $\hat I_1$. Repeating the same approach but setting $\hat I_1=\hat I_3=\dots=\hat I_{\hat N}=0$ one derives the coefficients $c_{2}(n_1,\dots,n_P)$ and so on. To obtain the complete solution of the IBP system one simply needs to add all $\hat N$ independently derived projections. 

To the best of our knowledge, this strategy is new and has not been applied before
\footnote{After this work was made public we learned that the latest update of the program {\tt Kira} offers the option of computing the coefficients of a subset of masters. Since the algorithm behind this is not described in \cite{Maierhoefer:2017hyi}, nor in the program's manual, we are unable to comment on how it compares with our approach.}.
It is easy to see why it leads to the correct solution of the IBP equations. Its correctness follows from the fact that each integral $I_i$ has an expansion in the set of masters $\hat I_m$, i.e. at each step the IBP equations can be rewritten as a homogeneous linear combination of all master integrals. Since the IBP equations are themselves linear and homogeneous in terms of the integrals $I_i$, one can see that the IBP equations never mix projections belonging to different master integrals. In essence, our proposal states that each of these projections can be computed in isolation from the others. 

The IBP solving strategy described here is independent of the approach used for solving the system of IBP equations. In practice, we will use the standard Laporta algorithm but one does not have to. In fact, we arrived at this idea while trying to find a way for solving the IBP system in closed form. We hope to return to this in a future work.

We have checked the correctness of our strategy in a number of non-trivial examples, such as the complete two-loop four-point amplitude (cross-checked with the program {\tt Reduze}~\cite{Studerus:2009ye,vonManteuffel:2012np}) and a number of two-loop five-point planar and non-planar cases as explained in detail in sec.~\ref{sec:results}.

At this point it will be beneficial to contrast our strategy to the usual way of solving IBP identities and to discuss the origin of increased efficiency. To this end we need to introduce the notion of a {\it sector} which is well-known in the IBP literature.

A sector is effectively a sub-topology indexed by $0$s and $1$s and defined by the position of a subset of propagators. For example, $\left[1, 1, 1, 0, \dots, 0\right]$ represents a sector. In the notation of eq.~(\ref{eq:I_def}) this sector contains all integrals $I(n_1,\dots,n_P)$ for which $n_{1,2,3}>0$ while $n_{4,\dots,P}\leq 0$. The number of different propagators that define a sector is called its {\it weight}. For example, the sector $\left[1, 1, 1, 0, \dots, 0\right]$ is of weight 3. A sector is called a zero-sector if all integrals that belong to it vanish. For the massless two-loop five-point amplitudes, all sectors with weight $<$ 3 are zero-sectors. Some sectors with weight $\geq 3$ are also zero-sectors.

Our strategy can lead to a more efficient solving of the IBP system for several reasons. First, once $\hat N-1$ masters are set to zero, many sectors become zero-sectors and thus do not need to be computed. In practice, this is a major simplification.

Second, setting masters to zero at the outset of the calculation simplifies the intermediate steps. The reason is that, taking the example of the Laporta algorithm, the IBP equations that will be solved first are generated from seeds that are in some sense close to the master integrals.
\footnote{Assuming that, as is usually the case, the masters are chosen with the help of the same ordering criterion that is used to generate the seeds for solving the IBP equations.}
In this way the information about vanishing masters is incorporated into the resulting IBP equations early on in the solving process. In large systems with many masters, our strategy could lead to a significant reduction in the size of the intermediate expressions. This, in turn, would reduce the computer memory requirement that is the limiting factor in solving large problems. 

Third, by solving for one master at a time one can parallelize the problem by computing several projections at the same time. The amount of parallelization achieved is only restricted by the available computer memory and CPU. One should keep in mind that, as we explain in sec.~\ref{sec:results}, the run-times for different masters can be vastly different.

\section{Results}\label{sec:results}

For definiteness, in this work we focus on the squared two-loop amplitude $M=\langle A^{(2)}\vert A^{(0)}\rangle$ for the process $q\bar q\to q'\bar q' g$. From the viewpoint of the IBPs it is representative of the other massless five-point two-loop amplitudes.

\begin{figure}[t]
\includegraphics[width=1.04\linewidth]{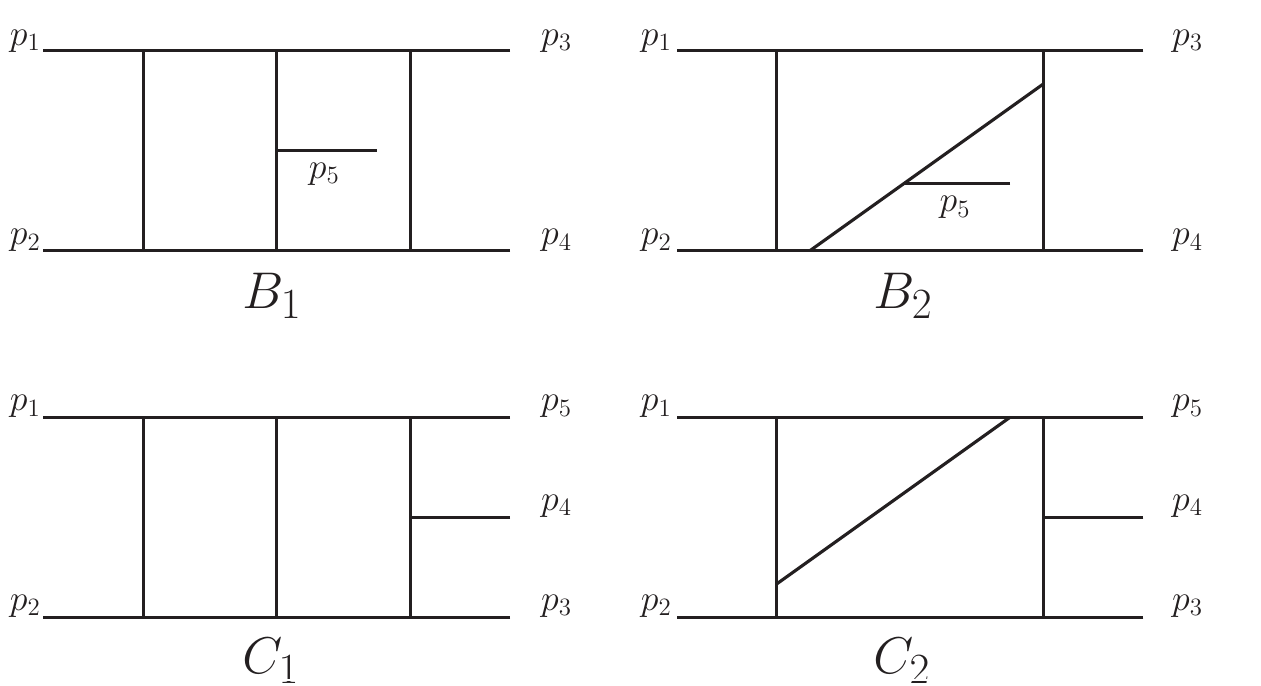}
\caption{The 8-propagator topologies $B_1, B_2, C_1$ and $C_2$. $B_1$ and $C_1$ are the most complicated non-planar and planar topologies, respectively.}
\label{fig:topologies}
\end{figure}

The Feynman integrals appearing in $M$ belong to several topologies. We label the family of non-planar ones $B$ and the family of planar ones $C$. There are two non-planar topologies ($B_1$ and $B_2$) that have the maximum possible number of propagators (eight) as well as two computationally simpler topologies with fewer than eight propagators. For the planar case, we have two topologies with eight propagators ($C_1$ and $C_2$) and one more, $C_3$, with seven propagators. All master integrals needed in the computation of the three planar $C$ topologies have been computed in analytic form \cite{Papadopoulos:2015jft} within the approach of ref.~\cite{Papadopoulos:2014lla}. Work towards the remaining non-planar ones is ongoing \cite{Chicherin:2017dob,Chicherin:2018ubl}. The four topologies with the maximum number of propagators are shown in fig.~\ref{fig:topologies}. 

The $B$ and $C$ families of topologies are defined through the following sets of 11 propagators:
\begin{eqnarray}
B&=&\left\{k_1^2, k_2^2, (k_1+p_1)^2, (k_1+p_1+p_2)^2, \right. \nonumber\\
&&\left. (k_2-p_3)^2, (k_2-k_1-p_3)^2, \right. \nonumber\\ 
&&\left. (k_2-k_1-p_1-p_2+p_4)^2, (k_2+p_4)^2, \right.\nonumber\\
&&\left. (k_2+p_1+p_2)^2, (k_2+p_1)^2, (k_1+p_3)^2\right\}  \label{eq:topologyB}\\
C&=&\left\{k_1^2, k_2^2, (k_1+p_1+p_2)^2, (k_1-k_2)^2, \right. \nonumber\\
&&\left. (k_2+p_1)^2, (k_2+p_1+p_2)^2, (k_2-p_3)^2, \right.\nonumber\\
&&\left. (k_1+p_1+p_2-p_3)^2, (k_1+p_1+p_2-p_3-p_4)^2,\right. \nonumber\\ 
&&\left. (k_2-p_3-p_4)^2, (k_1+p_1)^2\right\}\,. \label{eq:topologyC}
\end{eqnarray}
The momenta $p_1$ and $p_2$ are incoming while $p_3$ and $p_4$ are the two independent outgoing momenta. 

The four 8-propagator topologies shown in fig.~\ref{fig:topologies} as well as the 7-propagator one, $C_3$, are defined by their highest-weight sectors (see sec.~\ref{sec:approach} for definitions)
\begin{eqnarray}
B_1&=&B\left[1, 1, 1, 1, 1, 1, 1, 1, 0, 0, 0\right]\,,\nonumber\\
B_2&=&B\left[1, 1, 1, 1, 0, 1, 1, 1, 0, 0, 1\right]\,,\nonumber\\
C_1&=&C\left[1, 1, 1, 1, 1, 1, 0, 1, 1, 0, 0\right]\,,\nonumber\\
C_2&=&C\left[1, 1, 1, 1, 1, 0, 0, 1, 1, 0, 1\right]\,,\nonumber\\
C_3&=&C\left[1, 0, 1, 1, 1, 1, 0, 1, 0, 0, 1\right]\,.
\label{eq:topologies}
\end{eqnarray}

We have identified the master integrals in each of the five topologies in eq.~(\ref{eq:topologies}). We find 113 masters in $B_1$, 75 in $B_2$, 62 in $C_1$, 28 in $C_2$ and 10 in $C_3$. Their explicit definitions, in the notation of eqs.~(\ref{eq:topologyB},\ref{eq:topologyC},\ref{eq:topologies}), can be found in an electronic file attached to this paper.

In this work we have computed and are making publicly available all coefficients $c_{i,m}$ belonging to the most complicated planar topology $C_{1}$ needed for the evaluation of the amplitude $q\bar q\to q'\bar q' g$. This includes the results for all required integrals with irreducible numerators of power as high as -5 and/or squared denominators.

To demonstrate the power and flexibility of our strategy, we have also computed and present here the coefficients of the masters belonging to the highest-weight sector (with weight=8) for topologies $B_1$ (9 masters) and $B_2$ (3 masters). We have computed all integrals with numerator powers as high as -6 and/or a squared denominator. All results mentioned above are available for download in electronic form from the following website \cite{website}.

Our results have been cross-checked in the following ways: the masters for all five topologies in eq.~(\ref{eq:topologies}) have been independently derived with {\tt Reduze}~\cite{Studerus:2009ye,vonManteuffel:2012np}. Using the results in refs.~\cite{Gluza:2010ws,Kosower:2018obg} we have related all (five) integrals with irreducible numerators of power -5 belonging to topology $C_1$ to integrals with lower numerator powers. Using our calculation for those integrals with lower numerator powers we find complete agreement with our direct calculation of the integrals with numerators of power -5. We have checked that this agreement holds for the projections on to the full set of masters in topology $C_1$. This is a highly non-trivial check for both our calculation and the results in refs.~\cite{Gluza:2010ws,Kosower:2018obg}. 

We have also checked that our calculation for topology $B_2$ agrees with the results in ref.~\cite{Boehm:2018fpv} by comparing all integrals with numerator powers of -4 (which is the highest numerator power computed in that paper). Ref.~\cite{Boels:2018nrr} has claimed to compute the planar integrals with numerator power -5 with the help of the program {\tt FIRE} \cite{Smirnov:2008iw,Smirnov:2014hma}. However, since that reference does not provide explicit results or details about their calculation, we are unable to compare.

A few comments about our calculation are in order. We have implemented the strategy proposed in this work in a private {\tt C++} code. A bottleneck in solving the IBP identities is the manipulation of large rational expressions. To that end we have used the program {\tt Fermat} \cite{Fermat}. 

The run-times for different master integrals are vastly different. The calculation of the coefficients of the master integrals in the highest-weight sector (i.e. those with the maximum number of propagators) is simplest and takes only a few minutes. The calculations corresponding to masters with fewer propagators, however, become progressively more complex and can be orders of magnitude slower. 

The projections which are hardest to compute are the ones corresponding to the masters of lowest weight (the ones with 3 propagators for the two-loop five-point massless case). We have found that the difference between the run-times among the set of masters of lowest weight belonging to the same topology (there are six such masters in topology $C_1$) spans an order of magnitude. 

The solutions of the full set of IBP identities in compressed format are in excess of 20~GB and are available for download from the website~\cite{website}. We have not attempted to simplify the expressions for the individual coefficients $c_{i,m}$ since such a simplification is likely to be useful only at the level of the complete amplitude eq.~(\ref{eq:Mhat}).

When computing the squared amplitude $\langle A^{(2)}\vert A^{(0)}\rangle$ for the process $q\bar q\to q'\bar q' g$ we have used the program {\tt Reduze}~\cite{Studerus:2009ye,vonManteuffel:2012np} for the generation of the Feynman diagrams, for their squaring and for the summation over color and spin traces. Some of those calculations have been sped up with the help of the program {\tt FORM} \cite{Vermaseren:2000nd}. The Feynman diagrams have been visualized with the help of the program {\tt JaxoDraw} \cite{Binosi:2008ig}.

\section{Conclusions}

In this work we propose and develop a new strategy for solving the IBP identities. With its help we are able to solve in analytic form the complete set of IBP identities required for the construction of all planar two-loop five-point massless QCD amplitudes with quarks and/or gluons. Since all required planar master integrals are known, the problem of the derivation of the planar five-point two-loop amplitudes in QCD is thus solved. 

The gigabyte-size of the resulting expressions makes their numerical evaluation non-trivial. A dedicated effort will be required if one is to use them for collider phenomenology. We hope to report on a parallel effort in this direction in the near future.

With the completion of the planar amplitudes all attention now turns towards the remaining non-planar topologies, which constitute a much harder problem than the planar ones. Based on our experience in the context of the present work we believe that our strategy will be able to solve this problem in an acceptable timeframe.

\begin{acknowledgments}
We thank Zahari Kassabov for related collaboration and Simon Badger, Michal Czakon, Costas Papadopoulos and Chris Wever for discussions. We thank Robert Lewis for support with the program {\tt Fermat}. A.M. thanks the Department of Physics at Princeton University for hospitality during the completion of this work. This work is supported by the European Research Council Consolidator Grant ``NNLOforLHC2". A.M. is also supported by the UK STFC Grants No. ST/L002760/1 and No. ST/K004883/1. 
\end{acknowledgments}

\end{document}